\documentclass[11pt]{article}
\usepackage{moriond,epsfig}

\def\apj{ApJ.}
\def\nat{Nature}

\def\gtrsim{ \lower .75ex \hbox{$\sim$} \llap{\raise .27ex \hbox{$>$}} }
\def\lesssim{ \lower .75ex \hbox{$\sim$} \llap{\raise .27ex \hbox{$<$}} }

\def\be{\begin{equation}}
\def\ee{\end{equation}}
\def\bea{\begin{eqnarray}}
\def\eea{\end{eqnarray}}

\begin{document}
\vspace*{4cm}
\title{Inhomogeneity in the Supernova Remnants as a Natural Explanation of the PAMELA/ATIC Observations}

\author{Tsvi Piran$^1$, Nir J. Shaviv$^1$ and Ehud Nakar$^2$ }

\address{1. Racah Institute for Physics, The Hebrew University, Jerusalem 91904, Israel \\
2.  The Raymond and Beverly Sackler School of Physics \& Astronomy, \\ Tel-Aviv University, Tel-Aviv 69978, Israel
}

\maketitle\abstracts{
Recent measurements of the positron/electron ratio in the cosmic ray
(CR) flux exhibits an apparent anomaly\cite{PAMELA}, whereby this
ratio increases between 10 and 100 GeV. In contrast, this ratio
should decrease according to the standard scenario, in which CR positrons are
secondaries  formed by hadronic interactions between the primary CR
protons and the interstellar medium (ISM)\cite{Strong1998leptons}.
The positron excess is therefore interpreted as evidence for either an
annihilation/decay of weakly interacting massive
particles, or for a direct astrophysical
source of pairs. The
common feature of all proposed models is that they invoke new
physics or new astrophysical sources. However, this line of argumentation
relies implicitly on the assumption of a relatively
homogeneous CR source distribution. Inhomogeneity of CR
sources on a scale of order a kpc, can naturally explain this
anomaly. If the nearest major CR source is about a kpc away, then
low energy electrons ($\sim 1$~GeV) can easily reach us.  At
higher energies ($\gtrsim 10$~GeV), the source electrons cool via
synchrotron and inverse-Compton  before reaching the solar vicinity.
Pairs formed in the local vicinity through the proton/ISM
interactions can reach the solar system also at high energies, thus
increasing the positron/electron ratio. A natural origin of source
inhomogeneity is the strong concentration of supernovae to the
galactic spiral arms. Assuming supernova remnants (SNRs) as the sole
primary source of CRs, and taking into account their concentration
near the galactic spiral arms, we consistently predict the observed
positron fraction between 1 and 100 GeV, while abiding to different
constraints such as the observed electron spectrum and the CRs
cosmogenic age. An ATIC\cite{ATIC} like electron spectrum excess at $\sim
600$~GeV can be explained, in this picture,  as the
contribution of a few known nearby SNRs.
}

PAMELA\cite{PAMELA} discovered that  the CR positron/electron ratio
increases with energy above $\sim 7$~GeV. The apparent discrepancy
between the theoretical standard prediction of a decreasing ratio
and these measurements is now commonly known as the ``PAMELA
anomaly''\cite{PAMELA_Moriond}. It is commonly interpreted as
evidence for a new source of primary CR positrons, most likely WIMPs
or pulsars. Measurements of the electron spectrum at
$0.1-1$~TeV by ATIC\cite{ATIC} show an excess of CR electrons at
energies of $300-800$~GeV, and at even higher energies ($1-4$~TeV)
HESS measures\cite{HESS} a sharp decay in the electron spectrum.
ATIC's results are usually considered as support of a dark matter
origin for the PAMELA anomaly, where the observed spectral bump
corresponds to the WIMP mass. Note however that the recent Fermi
results\cite{Fermi}, exhibit a significantly smaller spectral excess
relative to standard CR diffusion models.

In the standard picture\cite{Strong1998leptons}, the majority of CRs are thought to
originate in SNR shocks.  SNRs,
however, are not expected to be a major  source of CR positrons.
Instead, as CR protons diffuse through the Galaxy, they collide with
interstellar medium (ISM) nuclei, producing
``secondary" positrons and electrons.
CRs diffuse within the disk, and escape the
Galaxy once they reach the halo height, $l_H \sim 1$~kpc above the
disk. The  diffusion coefficient can be approximated as $D = D_0
(E/E_0)^{\beta}$. Most CR diffusion models assume
that CRs are produced with a power-law spectrum, $N_E \equiv dN /dE
\propto E^{-\alpha}$.  The observed spectrum  is then a convolution
of the source spectrum and propagation losses, giving for the
primary electrons  $N_{E,obs}^{(e)}\propto
E^{-(\alpha_e+\beta)}$. Positrons are secondary CRs formed from CR
protons, and suffer additional propagation loses, implying
$N_{E,obs}^{(s)}\propto N_{E,obs}^{(p)}E^{-\beta}\propto
E^{-(\alpha_p+2\beta)}$.  The predicted flux ratio is
$\phi^+/(\phi^- + \phi^+) \approx \phi^+/\phi^- \propto
E^{\alpha_e-\alpha_p-\beta}$, where $\alpha_e$ and $\alpha_p$ are
the source power-law indices of electrons and protons respectively.
Both electrons and protons are expected to
have similar spectral slopes, i.e., $\alpha_e \approx \alpha_p$,
which is somewhat larger than 2.  Consequently, $\alpha_p-\alpha_e
< \beta \approx 0.3-0.6$ and the standard model predicts, in
contrast to PAMELA observations, a CR positron/electron ratio which
decreases with energy.

This standard model assumes a homogenous,  source
distribution\cite{Strong1998leptons,Strong1998nucleons}.
However, as star formation in
spiral galaxies is concentrated in
spiral arms\cite{LaceyDuric,ShavivNewAstronomy} one should
consider the effect of  inhomogeneities
in the CR source distribution on the CR spectrum. This
inhomogeneity of sources influences the electrons/positrons spectra
via cooling which sets a typical distance scale that an
electron/positron with a given energy can diffuse away from its
source. For a homogenous distribution cooling affects  the spectra
of (primary) electrons and (secondary) positrons in the same way and
their ratio is unaffected. On the other hand, primary electrons will
be strongly affected by an inhomogeneous source distribution at
energies for which the diffusion time is longer than the cooling
time. Protons are not affected by cooling and are therefore
distributed rather smoothly in the galaxy even if their sources are
inhomogeneous. The secondary positrons (that are produced by the
smoothly distributed protons) are only weakly affected by the
inhomogeneity of the sources. This effect would induce an observed
signature  on $\phi^+/\phi^-$, with similar properties to the one
observed by PAMELA.

We \cite{we} considered a simple analytic model for diffusion from
a source at a distance $d$  from Earth.
We model the
galaxy as a two dimensional
slab.  The  Galactic plane is
infinite and the  disk height is finite, $l_H$.
The source is at a distance $d$ from Earth. A CR diffuses
within this slab with a constant diffusion coefficient $D(E)$,  and
it escapes once $|y|>l_H$.
We find that for  a  a turnover in $\phi^+/\phi^-$ is
observed at $E_b$ which satisfies $\tau_c(E_b) \approx
\min\{\tau_x(E_b),(\tau_e(E_b) \tau_x(E_b))^{1/2}\}$.
$\phi^+/\phi^-$  for $E<E_b$ decreases, while it increases for $E>E_b$.
This is  the observed behavior seen by PAMELA, provided that $E_b \approx 10$~GeV, which
the case using typical parameters for cooling and diffusion from a
 source at  $d \approx 1$~kpc \cite{we}. The nearest spiral arm to
the solar system is the Sagittarius-Carina arm at a distance of
$\approx 1$~kpc.

At
the same time the typical age of CR protons with energy $E_b$ is $a
\sim \max\{\tau_e,(\tau_e\tau_d)^{1/2}\}$. Therefore a natural
prediction of the model is $a(E_b) \gtrsim \tau_c(E_b)$ and a
comparison of the two observables can be used as a consistency test
for the model. Moreover, over a wide range of the parameter space for
 which $d \gtrsim l_H$, the model predicts $a(E_b) \approx \tau_c(E_b)$
regardless of the value of the diffusion coefficient $D$.

To demonstrate  quantitatively the potential of this model to
recover the observed behavior of  $\phi^+/\phi^-$, we \cite{we} (see also ref.\
8\nocite{ShavivNewAstronomy}) simulated
numerically the CR diffusion  for  a realistic spiral-arm
concentrated source distribution. Before presenting these results we
stress that all other models explaining PAMELA  invoke a new ad hoc
source of high energy CR positrons which has  a  negligible effect
on low energy CR components. However, in our model, the PAMELA
explanation is intimately related to low and intermediate energy CR
propagation in the Galaxy. Namely, by revising the source
distribution of CRs, we affect numerous properties of $\sim$\ GeV
CRs. Given that the interpretation of observations (in particular,
isotopic ratios) used to infer model parameters (such as  $D_0$,
$\beta$ or $l_H$) depend on the complete model, one should proceed
while baring in mind that  these parameters may differ in our model
from present canonical values. In this sense, the objective
is not to carry a comprehensive parameter study, fitting the
whole CR data set to an inhomogeneous source distribution model.
Instead, our goal is to demonstrate the potential of the model to
explain naturally the PAMELA anomaly. To this end we use the
simplest possible model, fixing all parameters with the exception of
the halo size, $l_H$,  and the normalization of the  diffusion
coefficient, $D_0$, that we vary to fit the data.

Small scale inhomogeneities are important at energies larger than a
few hundreds GeV, for which the lifetime, and therefore propagation
distance, of electrons is so short that  the electron spectrum is
dominated by a single, or at most a few nearby
sources\cite{Atoyan,Kobayashi,Profumo}. To take this  effect into
account  we truncate  the ``homogeneous'' disk component  at
$r<0.5$~kpc and age less than $t<0.5$~Myr, and we add all
known SNRs within this 4-volume: Geminga, Monogem, Vela, Loop I and
the Cygnus Loop,  as discrete instantaneous sources. These sources
were described using the analytical solution\cite{Atoyan} for the
diffusion and cooling from an instantaneous point source.

\begin{figure}[tbh]
\centerline{\begin{minipage}{3.5in}
\epsfig{file=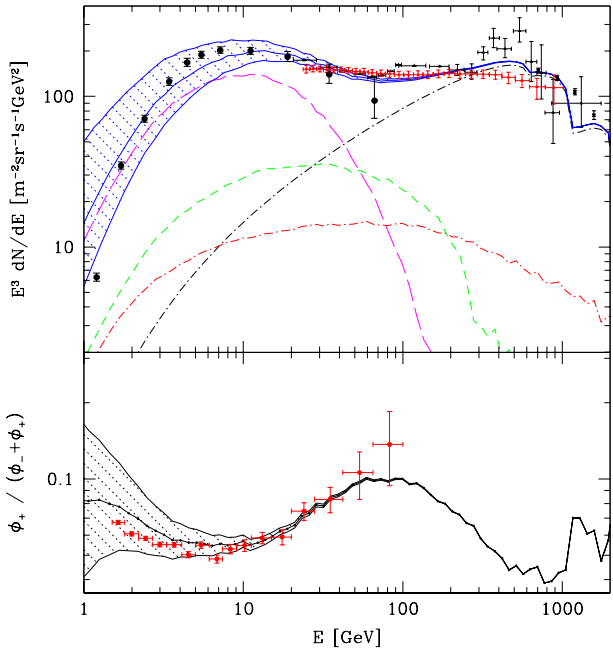,width=3.5in}
\end{minipage}
\begin{minipage}{2.2in}
\vskip -0.5cm
\caption{
{\it Bottom Panel}:
Model results and the measured PAMELA
points for the positron fraction.  The shaded region is the variability expected from solar
modulation effects$^{13}$. 
{\it Top Panel}: The
expected electron and positron spectra -- Primary arm electrons
(long dashed purple), primary disk electrons with nearby sources
excluded (short dashed green), nearby SNRs (dot-dashed black),
secondary positrons (dot-dashed red), and their sum (blue). The
hatched region describes the solar modulation range (from 200 MV to 1200 MV).
The four data sets
plotted are of HEAT$^{15}$ (circles), ATIC$^{3}$ 
(triangles), HESS$^{5}$ (open squares) and Fermi$^{6}$ (red triangles). 
}
\end{minipage}}
\label{fig:PressureFlux}
\end{figure}

The lower panel of fig.\ 1 depicts $\phi^+ / (\phi^+ + \phi^-)$.
 As expected from the simple analytical
model, the fraction decreases up to $\sim 10$ GeV and then it  starts
increasing.
At about 100 GeV, the ratio flattens and it decreases above this energy because of the
 injection of ``fresh" CRs from recent nearby SNRs whose high energy primary electrons don't have
time to cool.  These sources also contribute to higher energy electrons.
The cosmogenic age we obtain in this model for 1\ GeV per nucleon particles is 14 Myr.

The upper panel of fig.\ 1 depicts the electronic spectrum and its
constituents---primary spiral arm electrons, primary disk electrons
(without nearby sources), the spectrum of the nearby sources and the
secondary pairs. Evidently, there are two small humps in the $E^3 N_E$
plot. The lower energy hump arises from spiral arm electrons, the
higher energy of which cannot reach us due to cooling. At higher energies, the spectrum flattens out because of local SNR contribution.
 For our nominal CR injection per SNe\cite{we}, we obtain a spectrum laying between Fermi\cite{Fermi} and ATIC\cite{ATIC}, and which appears like a small hump. The three ``steps" in it are due to the cooling cutoffs from
Geminga, Loop I and the Monogem SNRs.
Note that the average CR flux
from these sources is about 3 to 6 times higher than can be expected
from the average disk population were it not truncated. This is not
surprising given that our local inter-arm region is perturbed by the
Orion Spur.

While the predictions for the positron/electron
ratio for the spiral
arms CR  model are very different than for  a homogenous sources
distribution, the effect on the electron spectrum is much more
subtle. Both models predict a break of the electron spectrum at $10$
GeV. The break predicted by spiral arm  model is from a
power law to an exponential, while in the homogenous
model it is a broken power-law. Given that above $\sim 100$
GeV the electron spectrum is strongly affected by local sources, the energy range between 10 to 100 GeV is too short  to distinguish, based on the
electron spectrum alone, between the two
models. Thus, while both models can adequately reproduce the observed
electron spectrum (at least up to 100 GeV),  only the
inhomogeneous source model can explain the positron/electron ratio.

One of the interesting predictions of this model where both the
PAMLEA and the ATIC anomalies are explained as consequences of
propagation effects from SNRs, is that the positron fraction should
start dropping with energy at $\sim 100$~GeV, just above the present
PAMELA measurement. It should reach a minimum around the ``ATIC peak",
where it should start rising again. Whether or not it can go up to
about $50$\% at a few TeV depends on whether the CRs from very
recent SNe, the Cygnus Loop and Vela, could have reached us or not.
This critically depends on the exact diffusion coefficient.  Here it
is also worth pointing out that above a few TeV the secondaries must
be produced within the local bubble, implying that their
normalization should be ten times lower than for the lower energy
secondaries. These predictions are in contrast to the case where
spectral features at higher energies are due to a primary source of pairs, in which case the
positron fraction is expected to keep rising also at a
few hundreds GeV. With these predictions, it will be straightforward
in the future to distinguish between propagation induced
``anomalies", and real anomalies arising from primary pairs (in
particular, when PAMELA's observations will extend to higher
energies). Of course, it is possible that the excess at high energies is due to a
source of primary pairs, while the PAMELA anomaly is a result of
SNRs in the spiral arms, but then it would force us to abandon the
simplicity of the model, that the anomalies are all due to
propagation effects from a source distribution borne from the known
structure of the Milky Way.

\section*{Acknowledgments}
We thank Marc Kamionkowski, Re'em Sari and Vasiliki Pavlidou for
helpful discussions. The work was partially supported by the ISF
center for High Energy Astrophysics, an ISF grant (NJS), an IRG
grant (EN) an ERC excellence grant and the Schwartzman Chair (TP).

\end{document}